\documentclass[aps,pra,floatfix,showpacs,10pt,twocolumn]{revtex4}
\usepackage[dvips]{graphicx}
\usepackage{amsmath,amsfonts,amssymb,graphics,graphicx,epsfig,color,times,bbm}

\begin{document}

\title{Entanglement in the anisotropic Heisenberg XYZ model with different Dzyaloshinskii-Moriya interaction and inhomogeneous magnetic field}
\author{Da-Chuang Li$^{1,2}$\footnote{E-mail: dachuang@ahu.edu.cn}, Zhuo-Liang Cao$^{1,
2}$\footnote{E-mail:zhuoliangcao@gmail.com (Corresponding
Author)}}

\affiliation{$^{1}$Department of Physics, Hefei Teachers
College, Hefei 230061 P. R. China\\
$^{2}$School of Physics {\&} Material Science, Anhui University,
Hefei 230039 P. R. China}

\pacs{03.67.Mn, 03.65.Ud, 75.10.Jm}

\keywords{Thermal entanglement; Heisenberg XYZ model; DM coupling
interaction}

\begin{abstract}
We investigate the entanglement in a two-qubit Heisenberg XYZ
system with different Dzyaloshinskii-Moriya(DM) interaction and
inhomogeneous magnetic field. It is found that the control
parameters ($D_{x}$, $B_{x}$ and $b_{x}$) are remarkably different
with the common control parameters ($D_{z}$,$B_{z}$ and $b_{z}$)
in the entanglement and the critical temperature, and these
x-component parameters can increase the entanglement and the
critical temperature more efficiently. Furthermore, we show the
properties of these x-component parameters for the control of
entanglement. In the ground state, increasing $D_{x}$ (spin-orbit
coupling parameter) can decrease the critical value $b_{xc}$ and
increase the entanglement in the revival region, and adjusting
some parameters (increasing $b_{x}$ and $J$, decreasing $B_{x}$
and $\Delta$) can decrease the critical value $D_{xc}$ to enlarge
the revival region. In the thermal state, increasing $D_{x}$ can
increase the revival region and the entanglement in the revival
region (for $T$ or $b_{x}$), and enhance the critical value
$B_{xc}$ to make the region of high entanglement larger. Also, the
entanglement and the revival region will increase with the
decrease of $B_{x}$ (uniform magnetic field). In addition, small
$b_{x}$ (nonuniform magnetic field) has some similar properties to
$D_{x}$, and with the increase of $b_{x}$ the entanglement also
has a revival phenomenon, so that the entanglement can exist at
higher temperature for larger $b_{x}$.

\end{abstract}

\maketitle

\section{introduction}
Entanglement is one of the most fascinating features of quantum
mechanics and it provides a fundamental resource in quantum
information processing \cite{1}. As a simple system, Heisenberg
model is an ideal candidate for the generation and the
manipulation of entangled states. This model has been used to
simulate many physical systems, such as nuclear spins \cite{3},
quantum dots \cite{4}, superconductor \cite{5} and optical
lattices \cite{6}, and the Heisenberg interaction alone can be
used for quantum computation by suitable coding \cite{2}. In
recent years, the Heisenberg model, including Ising model
\cite{7}, XY model \cite{8}, XXX model \cite{9}, XXZ model
\cite{10} and XYZ model \cite{11,12}, have been intensively
studied, especially in thermal entanglement and quantum phase
transition. Very recently, F. Kheirandish \emph{et al.} and Z. N.
Gurkan \emph{et al.} \cite{12} discussed the Heisenberg XYZ model
with DM interaction (arising from spin-orbit coupling) and
magnetic field. However, almost all the directions of spin-orbit
coupling and the external magnetic field are fixed in the z-axis
in the above papers. The external magnetic field along other
directions is discussed only in some special Heisenberg models,
and the spin-orbit coupling parameter along other directions has
never been taken into account. This motivates us to think about
what different phenomena will appear if we choose the parameters
along different directions in the generalized Heisenberg XYZ
model. To explore this, here we choose x-axis as the direction of
spin-orbit coupling and the external magnetic field in the
Heisenberg XYZ model.

In this paper, we discuss the differences between the Heisenberg
XYZ models with parameters in different directions, and then
analyze the influences of these parameters on the ground-state
entanglement and thermal entanglement. We find that x-component DM
interaction and inhomogeneous magnetic field are more efficient
control parameters, and more entanglement and higher critical
temperature can be obtained by adjusting the value of these
parameters. In order to provide a detailed analytical and
numerical analysis, here we take concurrence as a measure of
entanglement \cite{13}. The concurrence $C$ ranges from 0 to 1,
$C=0$ and $C=1$ indicate the vanishing entanglement and the
maximal entanglement respectively. For a mixed state $\rho$, the
concurrence of the state is
$C(\rho)=\max\{2\lambda_{\max}-\sum_{i=1}^{4}\lambda_{i}, 0\}$,
where $\lambda_{i}s$ are the positive square roots of the
eigenvalues of the matrix
$R=\rho(\sigma^{y}\bigotimes\sigma^{y})\rho^{*}(\sigma^{y}\bigotimes\sigma^{y})$,
and the asterisk denotes the complex conjugate. If the state of a
system is a pure state, i.e. $\rho=|\Psi\rangle\langle\Psi|,
|\Psi\rangle=a|00\rangle+b|01\rangle+c|10\rangle+d|11\rangle$, the
concurrence will be reduced to $C(|\Psi\rangle)=2|ad-bc|$.

Our paper is organized as follows. In Sec. II, we compare the DM
coupling parameters along different directions. In Sec. III, we
compare the external magnetic fields (including uniform and
nonuniform magnetic field) along different directions. In Sec. IV,
we analyze the entanglement of generalized Heisenberg XYZ model
with x-component DM interaction and inhomogeneous magnetic field.
Finally, in Sec. V  a discussion concludes the paper.

\section{The comparison between the DM coupling parameters along different directions}
\subsection{DM coupling parameter along z-axis}

The Hamiltonian $H$ for a two-qubit anisotropic Heisenberg XYZ
model with z-component DM coupling and external magnetic field is
\begin{eqnarray}
\label{1}
H&=&J_{x}\sigma_{1}^{x}\sigma_{2}^{x}+J_{y}\sigma_{1}^{y}\sigma_{2}^{y}+J_{z}\sigma_{1}^{z}\sigma_{2}^{z}
+D_{z}(\sigma_{1}^{x}\sigma_{2}^{y}-\sigma_{1}^{y}\sigma_{2}^{x})\nonumber\\
&&+(B_{z}+b_{z})\sigma_{1}^{z}+(B_{z}-b_{z})\sigma_{2}^{z},
\end{eqnarray}
where $J_{i}(i=x,y,z)$ are the real coupling coefficients, $D_{z}$
is the z-component DM coupling parameter, $B_{z}$ (uniform
external magnetic field) and $b_{z}$ (nonuniform external magnetic
field) are the z-component magnetic field parameters,
$\sigma^{i}(i=x,y,z)$ are the Pauli matrices. The coupling
constants $J_{i}>0$ corresponds to the antiferromagnetic case, and
$J_{i}<0$ corresponds to the ferromagnetic case. This model can be
reduced to some special Heisenberg models by changing $J_{i}$.
Parameters $J_{i}, D_{z}, B_{z}$ and $b_{z}$ are dimensionless.

Using the similar process to \cite{12}, we can get the eigenstates
of $H$:
\begin{subequations}\label{2}
\begin{equation}
|\phi_{1,2}\rangle=\sin\theta_{1,2}|00\rangle+\cos\theta_{1,2}|11\rangle,
\end{equation}
\begin{equation}
|\phi_{3,4}\rangle=\sin\theta_{3,4}|01\rangle+\chi\cos\theta_{3,4}|10\rangle,
\end{equation}
\end{subequations}
with corresponding eigenvalues:
\begin{subequations}\label{3}
\begin{equation}
E_{1,2}=J_{z}\pm w_{1},
\end{equation}
\begin{equation}
E_{3,4}=-J_{z}\pm w_{2},
\end{equation}
\end{subequations}
where $w_{1}=\sqrt{4B^{2}_{z}+(J_{x}-J_{y})^{2}}$,
$w_{2}=\sqrt{4b^{2}_{z}+4D^{2}_{z}+(J_{x}+J_{y})^{2}}$, and
$\chi=\frac{J_{x}+J_{y}-2iD_{z}}{\sqrt{(J_{x}+J_{y})^2+4D^{2}_{z}}}$,
$\theta_{1,2}=\arctan(\frac{J_{x}-J_{y}}{\pm w_{1}-2B_{z}})$,
$\theta_{3,4}=\arctan(\frac{\sqrt{(J_{x}+J_{y})^2+4D^{2}_{z}}}{\pm
w_{2}-2b_{z}})$. The system state at thermal equilibrium (thermal
state) is $\rho(T)=\frac{\exp(\frac{-H}{K_{B}T})}{Z}$, where
$Z=Tr[\exp(\frac{-H}{K_{B}T})]$ is the partition function of the
system, $H$ is the system Hamiltonian, $T$ is the temperature and
$K_{B}$ is the Boltzmann costant which we take equal to 1 for
simplicity. Then we can get the analytical expressions of $Z,
\rho(T)$, and $R$, but we do not list them because of the
complexity. After straightforward calculations, the positive
square roots of the eigenvalues of $R$ can be expressed as:
\begin{subequations}\label{4}
\begin{eqnarray}
\lambda_{1,2}&=&\frac{e^{\frac{J_{z}}{T}}}{Zw_{2}}\bigg|\sqrt{w^{2}_{2}\cosh^{2}(\frac{w_{2}}{T})-4b^{2}_{z}\sinh^{2}(\frac{w_{2}}{T})}\nonumber\\
&&\pm\sinh(\frac{w_{2}}{T})\sqrt{(J_{x}+J_{y})^2+4D^{2}_{z}}\bigg|,
\end{eqnarray}
\begin{eqnarray}
\lambda_{3,4}&=&\frac{e^{\frac{-J_{z}}{T}}}{Zw_{1}}\bigg|\sqrt{w^{2}_{1}\cosh^{2}(\frac{w_{1}}{T})-4B^{2}_{z}\sinh^{2}(\frac{w_{1}}{T})}\nonumber\\
&&\pm\sinh(\frac{w_{1}}{T})(J_{x}-J_{y})\bigg|.
\end{eqnarray}
\end{subequations}
Thus, the concurrence of this system can be written as \cite{13}:
\begin{equation}\label{5}
C=\max\{|\lambda_{1}-\lambda_{3}|-\lambda_{2}-\lambda_{4},0\},
\end{equation}
Fig. 1(a) demonstrates the concurrence versus temperature $T$ and
$D_{z}$, where $J_{x}=1$, $J_{y}=0.5$, $J_{z}=0.2$, and
$B_{z}=b_{z}=0$. It shows that the concurrence will decrease with
increasing temperature $T$ and increase with increasing $D_{z}$
for a certain temperature. The critical temperature $T_{c}$ is
dependent on $D_{z}$, increasing $D_{z}$ can increase the critical
temperature above which the entanglement vanishes. These results
are in accord with those in Ref. \cite{12}.

\subsection{DM coupling parameter along x-axis}
The Hamiltonian $H^{'}$ for a two-qubit anisotropic Heisenberg XYZ
model with x-component DM coupling and external magnetic field is
\begin{eqnarray}
\label{6}
H^{'}&=&J_{x}\sigma_{1}^{x}\sigma_{2}^{x}+J_{y}\sigma_{1}^{y}\sigma_{2}^{y}+J_{z}\sigma_{1}^{z}\sigma_{2}^{z}
+D_{x}(\sigma_{1}^{y}\sigma_{2}^{z}-\sigma_{1}^{z}\sigma_{2}^{y})\nonumber\\
&&+(B_{x}+b_{x})\sigma_{1}^{x}+(B_{x}-b_{x})\sigma_{2}^{x},
\end{eqnarray}
similarly, where $J_{i}(i=x,y,z)$ are the real coupling
coefficients, $D_{x}$ is the x-component DM coupling parameter,
$B_{x}$ (uniform external magnetic field) and $b_{x}$ (nonuniform
external magnetic field) are the x-component magnetic field
parameters, and $\sigma^{i}(i=x,y,z)$ are the Pauli matrices. The
coupling constants $J_{i}>0$ corresponds to the antiferromagnetic
case, and $J_{i}<0$ corresponds to the ferromagnetic case. This
model can be reduced to some special Heisenberg models by changing
$J_{i}$. Here, all the parameters are dimensionless, and we
introduce the mean coupling parameter $J$ and the partial
anisotropic parameter $\Delta$ in the YZ-plane, where
$J=\frac{J_{y}+J_{z}}{2}$ and
$\Delta=\frac{J_{y}-J_{z}}{J_{y}+J_{z}}$.

\begin{figure}[tbp]
\includegraphics[angle=0,width=3.1in,height=1.5in,bb=36 423 555 665]{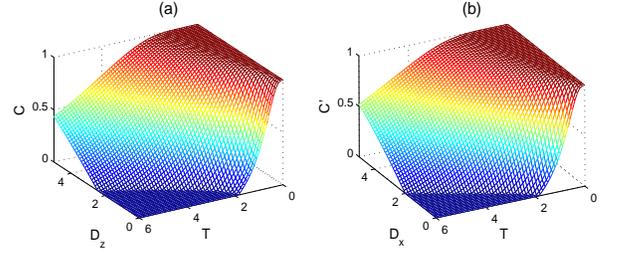}
\caption{(Color online) (a) Thermal concurrence versus $T$ and
$D_{z}$ without the external magnetic field. (b) Thermal
concurrence versus $T$ and $D_{x}$ without the external magnetic
field. Here $J_{x}=1, J_{y}=0.5$, and $J_{z}=0.2$.} \label{fig1}
\end{figure}
In the standard basis
$\{|00\rangle,|01\rangle,|10\rangle,|11\rangle\}$, the Hamiltonian
$H^{'}$ can be rewritten as:
\begin{equation}
\label{7} H^{'}=\left(
\begin{array}{cccc}
  J_{z} & G_{2} & G_{3} & J_{x}-J_{y} \\
  G_{4} & -J_{z} & J_{x}+J_{y} & G_{1} \\
  G_{1} & J_{x}+J_{y} & -J_{z} & G_{4} \\
  J_{x}-J_{y} & G_{3} & G_{2} & J_{z} \\
\end{array}
\right),
\end{equation}
where $G_{1,2}=iD_{x}+B_{x}\pm b_{x}, G_{3,4}=-iD_{x}+B_{x}\pm
b_{x}$. By calculating, we can obtain the eigenstates of $H^{'}$
\begin{subequations}\label{8}
\begin{eqnarray}
|\psi_{1,2}\rangle&=&\frac{1}{\sqrt{2}}(\sin\varphi_{1,2}|00\rangle+\cos\varphi_{1,2}|01\rangle\nonumber\\
&&+\cos\varphi_{1,2}|10\rangle+\sin\varphi_{1,2}|11\rangle),
\end{eqnarray}
\begin{eqnarray}
|\psi_{3,4}\rangle&=&\frac{1}{\sqrt{2}}(\sin\varphi_{3,4}|00\rangle+\chi^{'}\cos\varphi_{3,4}|01\rangle\nonumber\\
&&-\chi^{'}\cos\varphi_{3,4}|10\rangle-\sin\varphi_{3,4}|11\rangle),
\end{eqnarray}
\end{subequations}
with corresponding eigenvalues:
\begin{subequations}\label{9}
\begin{equation}
E^{'}_{1,2}=J_{x}\pm w^{'}_{1},
\end{equation}
\begin{equation}
E^{'}_{3,4}=-J_{x}\pm w^{'}_{2},
\end{equation}
\end{subequations}
where $\chi^{'}=\frac{-iD_{x}-b_{x}}{\sqrt{b^2_{x}+D^{2}_{x}}}$,
$w^{'}_{1}=\sqrt{4B^{2}_{x}+(J_{y}-J_{z})^{2}}$,
$w^{'}_{2}=\sqrt{4b^{2}_{x}+4D^{2}_{x}+(J_{y}+J_{z})^{2}}$, and
$\varphi_{1,2}=\arctan(\frac{2B_{x}}{J_{y}-J_{z}\pm w^{'}_{1}})$,
$\varphi_{3,4}=\arctan(\frac{2\sqrt{b^{2}_{x}+D^{2}_{x}}}{-J_{y}-J_{z}\pm
w^{'}_{2}})$. In the above standard basis, the density matrix
$\rho^{'}(T)$ has the following form:
\begin{equation}
\label{10} \rho^{'}(T)=\left(
\begin{array}{cccc}
  U_{1} & Q_{1}^{*} & Q_{2}^{*} & U_{2} \\
  Q_{1} & V_{1} & V_{2} & Q_{2} \\
  Q_{2} & V_{2} & V_{1} & Q_{1} \\
  U_{2} & Q_{2}^{*} & Q_{1}^{*} & U_{1} \\
\end{array}
\right),
\end{equation}
\begin{figure}[tbp]
\includegraphics[scale=0.58,angle=0]{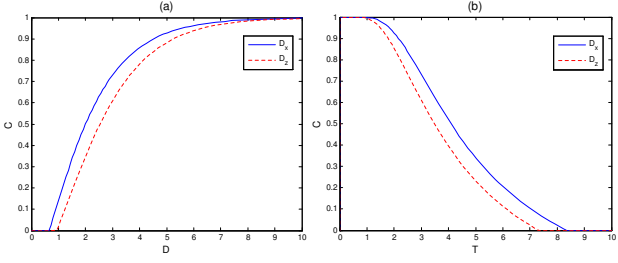}
\caption{(Color online) (a) Thermal concurrence versus DM coupling
parameter $D_{z}$ (red dotted line), $D_{x}$ (blue solid line),
where $T=3$. (b) Thermal concurrence versus temperature $T$ for
$D_{z}=3$ (red dotted line) and $D_{x}=3$ (blue solid line). Here
$J_{x}=1, J_{y}=0.5, J_{z}=0.2$, and $B_{z,x}=b_{z,x}=0$.}
\label{fig2}
\end{figure}
where
\begin{eqnarray*}
U_{1,2}&=&\frac{1}{2Z^{'}}(e^{-\frac{J_{x}+w_{1}^{'}}{T}}\sin^{2}\varphi_{1}+e^{-\frac{J_{x}-w_{1}^{'}}{T}}\sin^{2}\varphi_{2}\nonumber\\
&&\pm e^{\frac{J_{x}-w_{2}^{'}}{T}}\sin^{2}\varphi_{3}\pm
e^{\frac{J_{x}+w_{2}^{'}}{T}}\sin^{2}\varphi_{4}),
\end{eqnarray*}
\begin{eqnarray*}
V_{1,2}&=&\frac{1}{2Z^{'}}(e^{-\frac{J_{x}+w_{1}^{'}}{T}}\cos^{2}\varphi_{1}+e^{-\frac{J_{x}-w_{1}^{'}}{T}}\cos^{2}\varphi_{2}\nonumber\\
&&\pm e^{\frac{J_{x}-w_{2}^{'}}{T}}\cos^{2}\varphi_{3}\pm
e^{\frac{J_{x}+w_{2}^{'}}{T}}\cos^{2}\varphi_{4}),
\end{eqnarray*}
\begin{eqnarray*}
Q_{1,2}&=&\frac{1}{2Z^{'}}(e^{-\frac{J_{x}+w_{1}^{'}}{T}}\sin\varphi_{1}\cos\varphi_{1}+e^{-\frac{J_{x}-w_{1}^{'}}{T}}\sin\varphi_{2}\cos\varphi_{2}\nonumber\\
&&\pm
e^{\frac{J_{x}-w_{2}^{'}}{T}}\chi^{'}\sin\varphi_{3}\cos\varphi_{3}\pm
e^{\frac{J_{x}+w_{2}^{'}}{T}}\chi^{'}\sin\varphi_{4}\cos\varphi_{4}).
\end{eqnarray*}
Then the positive square roots of the eigenvalues of the matrix
$R^{'}$ can be obtained
\begin{widetext}
\begin{subequations}\label{11}
\begin{eqnarray}
\lambda^{'}_{1,2}=\frac{e^{\frac{J_{x}}{T}}}{Z^{'}}\bigg|\cosh(\frac{2w^{'}_{2}}{T})-\frac{8b_{x}^{2}}{w_{2}^{'2}}\sinh^{2}(\frac{w^{'}_{2}}{T})\pm\sqrt{\frac{4D^{2}_{x}+(J_{y}+J_{z})^{2}}{w_{2}^{'2}}\bigg[\sinh^{2}(\frac{2w^{'}_{2}}{T})-\frac{16b^{2}_{x}}{w_{2}^{'2}}\sinh^{4}(\frac{w_{2}^{'}}{T})\bigg]}\bigg|^{\frac{1}{2}},
\end{eqnarray}
\begin{eqnarray}
\lambda^{'}_{3,4}=\frac{e^{\frac{-J_{x}}{T}}}{Z^{'}}\bigg|\cosh(\frac{2w^{'}_{1}}{T})-\frac{8B_{x}^{2}}{w_{1}^{'2}}\sinh^{2}(\frac{w^{'}_{1}}{T})\pm\sqrt{\frac{(J_{y}-J_{z})^{2}}{w_{1}^{'2}}\bigg[\sinh^{2}(\frac{2w^{'}_{1}}{T})-\frac{16B^{2}_{x}}{w_{1}^{'2}}\sinh^{4}(\frac{w_{1}^{'}}{T})\bigg]}\bigg|^{\frac{1}{2}},
\end{eqnarray}
\end{subequations}
\end{widetext} where
$Z^{'}=2[e^{\frac{-J_{x}}{T}}\cosh(\frac{w_{1}^{'}}{T})+e^{\frac{J_{x}}{T}}\cosh(\frac{w_{2}^{'}}{T})]$.
Thus, according to \cite{13}, the corresponding concurrence of
this system can be expressed as:
\begin{equation}\label{12}
C^{'}=\max\{|\lambda^{'}_{1}-\lambda^{'}_{3}|-\lambda^{'}_{2}-\lambda^{'}_{4},0\},
\end{equation}
In Fig. 1(b), we plot the concurrence $C^{'}$ as a function of $T$
and $D_{x}$, where $J_{x}=1$, $J_{y}=0.5$, $J_{z}=0.2$, and
$B_{x}=b_{x}=0$. It shows the similar results to those in the last
subsection, but if we compare the two figures in Fig. 1 in detail,
we can easily find some differences between them. i.e., there is
less disentanglement region in Fig. 1(b) than in Fig. 1(a), and
increasing x-component parameter $D_{x}$ can make the entanglement
increase more rapidly, for example, when $T=6$ the concurrence
increases more rapidly in Fig. 1(b) than in Fig. 1(a). These
phenomena show that x-component parameter $D_{x}$ has some more
remarkable influences than the z-component parameter $D_{z}$ on
the thermal entanglement and the critical temperature. To
demonstrate the differences more clearly, we plot Fig. 2, which
shows the advantages of using parameter $D_{x}$. In Fig. 2(a), we
can easily find that for a certain temperature, $D_{x}$ has more
entanglement than $D_{z}$, and increasing $D_{x}$ can increase the
entanglement more rapidly. Also, in Fig. 2(b), we can see that for
the same $D_{x}$ and $D_{z}$, $D_{x}$ has a higher critical
temperature. In a word, increasing $D_{x}$ can enhance the
entanglement, and slow down the decrease of the entanglement (by
increasing critical temperature) more efficiently than $D_{z}$. So
by changing the direction of DM coupling interaction, we can get a
more efficient control parameter.

\section{The comparison between the external magnetic fields along different directions}
\subsection{External magnetic field along z-axis}
Here, we use the same model expressed in Eq. (1). According to Eq.
(4), we can write the expression of concurrence, and then analyze
the variation of the entanglement and the roles of the external
magnetic field along z-axis, by fixing other parameters.

In Fig. 3(a), the thermal concurrence as a function of the uniform
magnetic field is plotted, where $T=0.1$. It is shown that, for
small $|B_{z}|$, the concurrence has a high entanglement and does
not vary with the variation of $|B_{z}|$, and with increasing
$|B_{z}|$, the entanglement drops to zero suddenly at the critical
value of $|B_{z}|$, and then undergoes a revival before decreasing
to zero. In Fig. 4(a), the thermal concurrence is plotted versus
the nonuniform magnetic field with $T=0.2$. It is evident that
increasing $|b_{z}|$ will decrease the entanglement. In Fig. 3(b)
and Fig. 4(b), the concurrence as a function of the temperature
$T$ is plotted for the uniform magnetic field and the nonuniform
magnetic field, respectively. It is shown that the concurrence
decreases with increasing $T$ for $B_{z}=1$ (Fig. 3(b)) and
$b_{z}=1.5$ (Fig. 4(b)).
\begin{figure}[tbp]
\includegraphics[scale=0.58,angle=0]{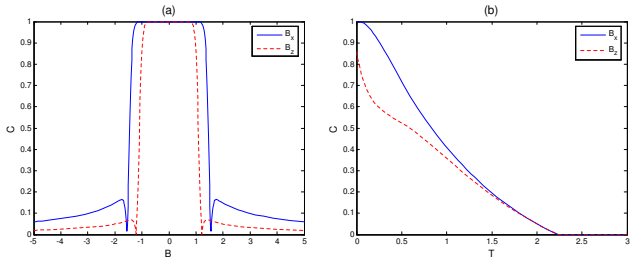}
\caption{(Color online) (a) Thermal concurrence versus uniform
magnetic field $B_{z}$ (red dotted line), $B_{x}$ (blue solid
line), where $T=0.1$. (b) Thermal concurrence versus temperature
$T$ for $B_{z}=1$ (red dotted line) and $B_{x}=1$ (blue solid
line). Here $J_{x}=1, J_{y}=0.8, J_{z}=0.2$, and
$D_{z,x}=b_{z,x}=0$.} \label{fig3}
\end{figure}

\begin{figure}[tbp]
\includegraphics[scale=0.58,angle=0]{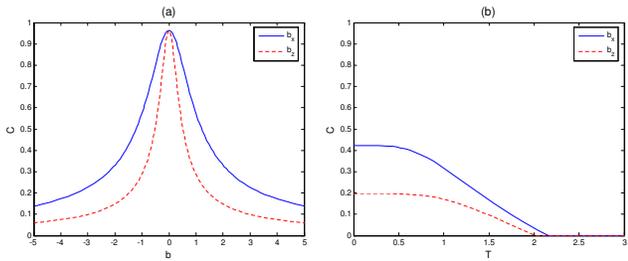}
\caption{(Color online) (a) Thermal concurrence versus nonuniform
magnetic field $b_{z}$ (red dotted line), $b_{x}$ (blue solid
line), where $T=0.2$. (b) Thermal concurrence versus temperature
$T$ for $b_{z}=1.5$ (red dotted line) and $b_{x}=1.5$ (blue solid
line). Here $J_{x}=0.2, J_{y}=0.4, J_{z}=1$, and
$D_{z,x}=B_{z,x}=0$.} \label{fig4}
\end{figure}

\subsection{External magnetic field along x-axis}
In this case, we use the model in Eq. (6). According Eq. (12), we
analyze similarly the variation of the entanglement and the roles
of the external magnetic field along x-axis, by fixing other
parameters.

In Fig. 3(a), by comparing with $|B_{z}|$, we can see that, though
$|B_{x}|$ has some similar properties to $|B_{z}|$, $|B_{x}|$ has
a larger critical value and a more remarkable revival. In Fig.
4(a), we find that the entanglement decreases more slowly with the
increase of nonuniform magnetic field $|b_{x}|$ than $|b_{z}|$,
and $|b_{x}|$ has more entanglement when $|b_{z}|=|b_{x}|$. Also,
In Fig. 3(b), we find that $B_{x}$ has more entanglement than
$B_{z}$ for a certain temperature when $B_{z}=B_{x}$. Fig. 4(b)
shows that, for the same $b_{x}$ and $b_{z}$, $b_{x}$ has a higher
critical temperature and more entanglement (for a certain
temperature).

It is evident that for some fixed parameters, using x-component
magnetic field can increase the entanglement more efficiently than
z-component magnetic field. So we can change the direction of the
external magnetic field to obtain a more efficient control
parameter.

\section{Heisenberg XYZ model with x-component DM interaction and inhomogeneous magnetic field}
According to Secs. II and III, we know that x-component DM
interaction and inhomogeneous magnetic field are more efficient
control parameters of entanglement. So in this section, we will
analyze the properties of parameters of the model (Eq. (6)) in
detail.

\subsection{Ground state entanglement}
When $T=0$, this system is in its ground state. It is easy to find
that the ground-state energy is equal to
\begin{subequations}
\begin{eqnarray}
E^{'}_{2}=J_{x}-w^{'}_{1},\,\,\,\,\,\,\,\,\,\,\,\,if
J_{x}<\frac{1}{2}(w_{1}^{'}-w_{2}^{'}),
\end{eqnarray}
\begin{eqnarray}
E^{'}_{4}=-J_{x}-w^{'}_{2},\,\,\,\,\,\,\,\,\,if
J_{x}>\frac{1}{2}(w_{1}^{'}-w_{2}^{'}),
\end{eqnarray}
\end{subequations}
and for $E^{'}_{2}$ and $E^{'}_{4}$, the ground state is
$|\psi_{2}\rangle$ and $|\psi_{4}\rangle$ respectively. At the
critical point $J_{x}=\frac{1}{2}(w_{1}^{'}-w_{2}^{'})$ (i.e.
$E^{'}_{2}=E^{'}_{4}$), the ground state is
$|G\rangle=\frac{1}{\sqrt{2}}(|\psi_{2}\rangle+|\psi_{4}\rangle)$
(an equal superposition of $|\psi_{2}\rangle$ and
$|\psi_{4}\rangle$). Thus, the concurrence of ground state can be
expressed as
\begin{widetext}
\begin{eqnarray}
\label{13} C^{'}(T=0)=\left\{
\begin{array}{l}
\bigg|\frac{J_{y}-J_{z}}{w_{1}^{'}}\bigg|,\,\,\,\,\,\,\,\,\,\,\,\,\,\,\,\,\,\,\,\,\,\,\,\,\,\,\,\,\,\,\,\,\,\,\,\,\,\,\,\,\,\,\,\,\,\,\,\,\,\,\,\,\,\,\,\,\,\,\,\,\,\,\,\,\,\,\,\,\,\,\,\,\,\,\,\,\,\,\,\,\,\,\,\,\,\,\,\,\,\,\,\,\,\,\,\,\,\,\,\,\,\,\,\,\,\,\,\,\,\,\,\,\,\,\,\,\,\,\,\,\,\,\,if\,\,\ J_{x}<\frac{1}{2}(w_{1}^{'}-w_{2}^{'}),\\
\frac{1}{2}\bigg|(\frac{J_{y}-J_{z}}{w_{1}^{'}}+\frac{J_{y}+J_{z}}{w_{2}^{'}})^{2}+\frac{4D_{x}^{2}}{w_{2}^{'2}}-\frac{2D_{x}^{2}(J_{y}-J_{z})(w_{2}^{'}+J_{y}+J_{z})}{(b_{x}^{2}+D_{x}^{2})w_{1}^{'}w_{2}^{'}}\bigg|^{\frac{1}{2}},\,\,\,\,\,if\,\,\ J_{x}=\frac{1}{2}(w_{1}^{'}-w_{2}^{'}),\\
\frac{\big|4D_{x}^{2}+(J_{y}+J_{z})^{2}\big|^{\frac{1}{2}}}{w_{2}^{'}},\,\,\,\,\,\,\,\,\,\,\,\,\,\,\,\,\,\,\,\,\,\,\,\,\,\,\,\,\,\,\,\,\,\,\,\,\,\,\,\,\,\,\,\,\,\,\,\,\,\,\,\,\,\,\,\,\,\,\,\,\,\,\,\,\,\,\,\,\,\,\,\,\,\,\,\,\,\,\,\,\,\,\,\,\,\,\,\,\,\,\,\,\,\,\,\,\,\,\,\,\,if\,\,\ J_{x}>\frac{1}{2}(w_{1}^{'}-w_{2}^{'}).\\
\end{array}
\right.
\end{eqnarray}
\end{widetext}

\begin{figure}[tbp]
\includegraphics[scale=0.5,angle=0]{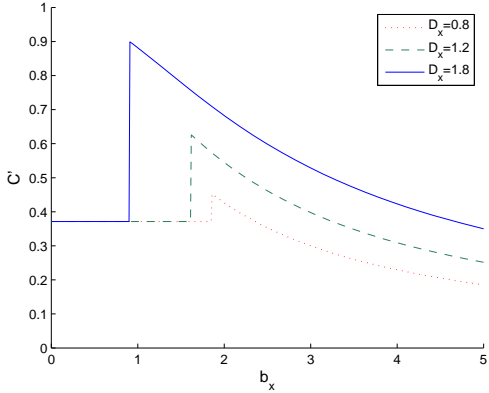}
\caption{(Color online) The ground-state concurrence versus
$b_{x}$ for different $D_{x}$. Here $J=0.5, \Delta=0.8, J_{x}=-1$,
and $B_{x}=1$.} \label{fig5}
\end{figure}

\begin{figure}[tbp]
\includegraphics[scale=0.46,angle=0]{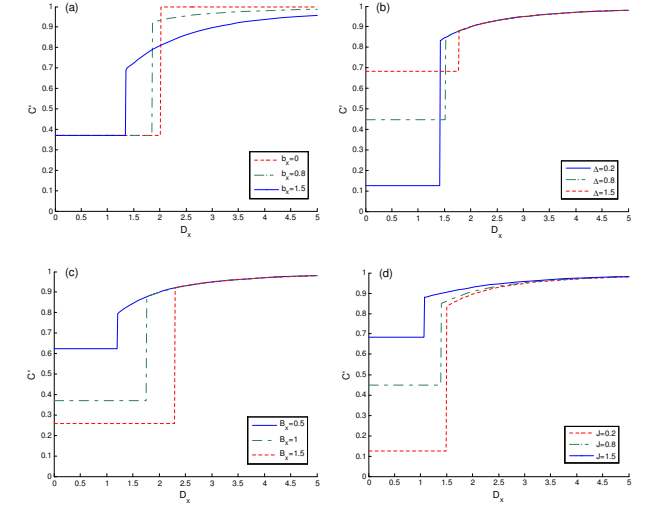}
\caption{(Color online) The ground-state concurrence versus
$D_{x}$ for different parameters. (a) $b_{x}=0$ (red dotted line),
0.8 (green Dash-dot line), and 1.5 (blue solid line), where
$J=0.5, \Delta=0.8, J_{x}=-1$, and $B_{x}=1$. (b) $\Delta=0.2$
(blue solid line), 0.8 (green Dash-dot line), and 1.5 (red dotted
line), where $J=0.5, J_{x}=-1,B_{x}=0.8$, and $b_{x}=1$. (c)
$B_{x}=0.5$ (blue solid line), 1 (green Dash-dot line), and 1.5
(red dotted line), where $J=0.5, \Delta=0.8, J_{x}=-1$, and
$b_{x}=1$. (d) $J=0.2$ (red dotted line), 0.8 (green Dash-dot
line), and 1.5 (blue solid line), where $\Delta=0.5,
J_{x}=-1,B_{x}=0.8$, and $b_{x}=1$.} \label{fig6}
\end{figure}
In Fig. 5, the ground-state concurrence is plotted as a function
of $b_{x}$ for different $D_{x}$. With increasing $b_{x}$, the
ground-state concurrence $C^{'}$ keeps a constant initially, and
then varies suddenly at the critical value of $b_{x}$
\big($b_{xc}=\frac{1}{2}\sqrt{(2J_{x}-w_{1}^{'})^{2}-(J_{y}+J_{z})^{2}-4D_{x}^{2}}$\big),
at which the quantum phase transition occurs. In the region of
$b_{x}>b_{xc}$, $C^{'}$ has a revival before decreasing to zero.
By increasing $D_{x}$, the critical value $b_{xc}$ will decrease,
and thus the revival region will become larger. In addition, there
will be more entanglement in the revival region for larger
$D_{x}$. The ground-state concurrence as a function of $D_{x}$ for
different parameters is shown in Figs. 6(a)-6(d). These figures
show that, with the increasing of $D_{x}$, the concurrence keeps a
constant initially, which depends on $J, \Delta$ and $B_{x}$,
before reaching the critical value
$D_{xc}=\frac{1}{2}\sqrt{(2J_{x}-w_{1}^{'})^{2}-(J_{y}+J_{z})^{2}-4b_{x}^{2}}$,
then varies suddenly at $D_{xc}$, and then undergoes a revival to
reach its maximum value ($C^{'}=1$) finally in the region of
$D_{x}>D_{xc}$. The critical point $D_{xc}$ decreases as $b_{x}$
and $J$ increase, and increases as $B_{x}$ and $\Delta$ increase.
Thus, by adjusting the parameters of the system (increasing
$b_{x}$ and $J$, decreasing $B_{x}$ and $\Delta$), we can decrease
the critical point $D_{xc}$ to get larger revival region with more
entanglement.

\subsection{Thermal entanglement}
As the temperature increases, the thermal fluctuation is
introduced into the system, thus the entanglement will be changed
due to the mix of the ground state and excited states. To see the
variation the entanglement, we use the thermal state $\rho^{'}(T)$
to describe the system state, and the concurrence $C^{'}$ to
measure the entanglement. According to Eq. (12), the concurrence
$C^{'}$ is plotted in Fig. 7 and Fig. 8 by fixing some parameters.

\begin{figure}[tbp]
\includegraphics[angle=0,width=3.2in,height=3in,bb=1 170 588 670]{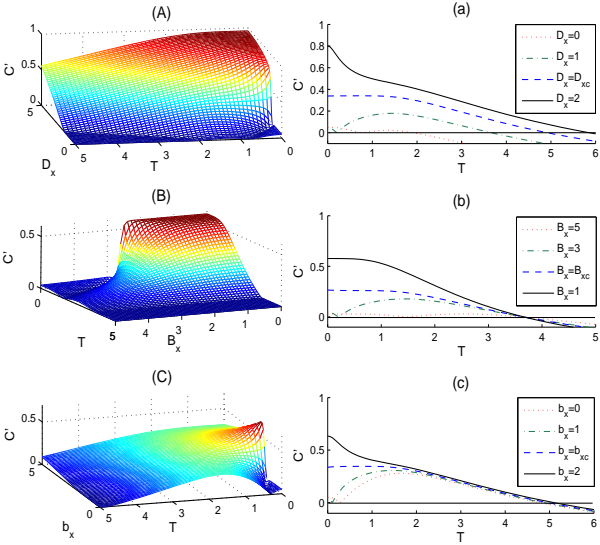}
\caption{(Color online) (A) Thermal concurrence versus $D_{x}$ and
$T$, where $B_{x}=3, b_{x}=1.5$. (a) Thermal concurrence versus
$T$ for $D_{x}=0$ (red dotted line), 1 (green Dash-dot line),
$D_{xc}\simeq1.575$ (blue dashed line), 2 (black solid line). (B)
Thermal concurrence versus $B_{x}$ and $T$, where $b_{x}=1.5,
D_{x}=1$. (b) Thermal concurrence versus $T$ for $B_{x}=5$ (red
dotted line), 3 (green Dash-dot line), $B_{xc}\simeq2.63$ (blue
dashed line), 1 (black solid line). (C) Thermal concurrence versus
$b_{x}$ and $T$, where $B_{x}=3, D_{x}=1.6$. (c) Thermal
concurrence versus $T$ for $b_{x}=0$ (red dotted line), 1 (green
Dash-dot line), $b_{xc}\simeq1.47$ (blue dashed line), 2 (black
solid line). Here $J_{x}=0.8, J_{y}=0.5$, and $J_{z}=0.2$.}
\label{fig7}
\end{figure}
Fig. 7 demonstrates the influence of some parameters on the
entanglement and critical temperature. In Fig. 7(A) and 7(a), when
$D_{x}$ is small, with increasing temperature the concurrence
decreases to zero at $T_{c1}$ (the first critical temperature),
and then undergoes a revival before decreasing to zero again at
$T_{c2}$ (the second critical temperature). When $D_{x}>D_{xc}$,
with increasing temperature the concurrence decreases to zero at
$T_{c2}$. Increasing $D_{x}$ can increase the revival region
(decreasing $T_{c1}$ and increasing $T_{c2}$) and enhance the
entanglement in the revival region, so the entanglement can exist
at higher temperature for larger $D_{x}$. In Fig. 7(B) and 7(b),
we see that decreasing $B_{x}$ can enhance the entanglement and
increase the revival region (decreasing $T_{c1}$), so the
entanglement has its maximum value when $B_{x}=0$ and $T=0$. Fig.
7(C) and 7(c) show that small $b_{x}$ has some similar properties
to $D_{x}$. In addition, Fig. 7(C) shows that the concurrence has
a revival phenomenon as $b_{x}$ increases, so the entanglement can
also exist at higher temperature for larger $b_{x}$.

On the whole, Fig. 7 can be divided into two regions: (i) The
region with revival. In this region, $D_{x}<D_{xc}, b_{x}<b_{xc}$,
or $B_{x}>B_{xc}$, where $D_{xc}, b_{xc}$ and $B_{xc}$ are the
critical values of the parameters. When $T<T_{c1}$,
$\lambda^{'}_{\max}=\lambda^{'}_{3}$ and thus
$C^{'}=\max\{\lambda^{'}_{3}-\lambda^{'}_{1}-\lambda^{'}_{2}-\lambda^{'}_{4},0\}$.
When $T=T_{c1}$, $\lambda^{'}_{3}=\lambda^{'}_{1}$ and thus
$C^{'}=0$. When $T>T_{c1}$, $\lambda^{'}_{\max}=\lambda^{'}_{1}$
and thus
$C^{'}=\max\{\lambda^{'}_{1}-\lambda^{'}_{3}-\lambda^{'}_{2}-\lambda^{'}_{4},0\}$.
With the increase of temperature, the entanglement decreases to
zero at $T_{c2}$ ultimately. (ii) The region without revival. In
this region, $D_{x}>D_{xc}, b_{x}>b_{xc}$, or $B_{x}<B_{xc}$. For
arbitrary temperature, $\lambda^{'}_{\max}=\lambda^{'}_{1}$, so
the concurrence is
$C^{'}=\max\{\lambda^{'}_{1}-\lambda^{'}_{3}-\lambda^{'}_{2}-\lambda^{'}_{4},0\}$.
Also, the entanglement vanishes finally at $T_{c2}$ with
increasing temperature.

\begin{figure}[tbp]
\includegraphics[width=3.1in,height=2.6in,angle=0,bb=5 176 590 690]{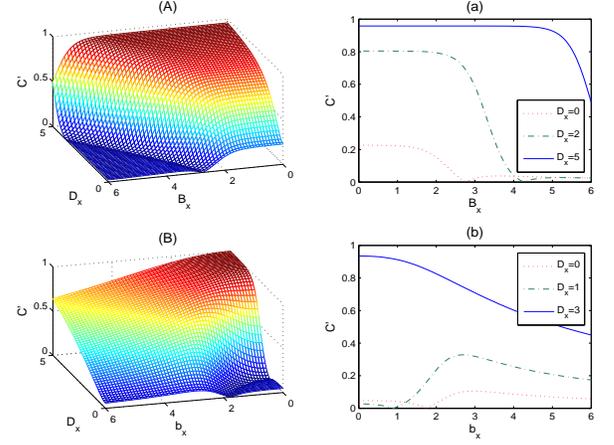}
\caption{(Color online) (A) Thermal concurrence versus $B_{x}$ and
$D_{x}$, where $b_{x}=1.5, T=0.5$. (a) Thermal concurrence versus
$B_{x}$ for $D_{x}=0$ (red dotted line), 2 (green Dash-dot line),
5 (blue solid line). (B) Thermal concurrence versus $b_{x}$ and
$D_{x}$, where $B_{x}=3, T=0.5$. (b) Thermal concurrence versus
$b_{x}$ for $D_{x}=0$ (red dotted line), 1 (green Dash-dot line),
3 (blue solid line). Here $J_{x}=0.8, J_{y}=0.5$, and
$J_{z}=0.2$.} \label{fig8}
\end{figure}

Fig. 8 shows the cross influence of the DM parameter and the
magnetic field on entanglement. In Fig. 8(A) and 8(a), when
$D_{x}$ is fixed, with increasing $B_{x}$ the entanglement is a
constant initially, and then dropped to zero suddenly at $B_{xc}$.
For $B_{x}>B_{xc}$, a revival phenomenon occurs, and the
entanglement becomes another constant. In addition, as $D_{x}$
increases, the critical value $B_{xc}$ increases, and the
entanglement enhances in the region of $B_{x}<B_{xc}$. In Fig.
8(B) and 8(b), when $D_{x}$ is small, with increasing $b_{x}$ the
entanglement is also a constant initially, then dropped to zero
suddenly at $b_{xc}$, and then undergoes a revival. Furthermore,
increasing $D_{x}$ can increase the revival region (decreasing
$b_{xc}$), and enhance the entanglement in the revival region.
When $D_{x}$ is large enough, with increasing $b_{x}$ the
entanglement decreases from its maximum value to zero.

\section{Discussion}
We have investigated the entanglement in the generalized two-qubit
Heisenberg XYZ system with different DM interaction and
inhomogeneous magnetic field. By comparing with the common
z-component parameters ($D_{z}$,$B_{z}$ and $b_{z}$), we find that
the x-component DM interaction and inhomogeneous magnetic field
($D_{x}$,$B_{x}$ and $b_{x}$) are more efficient control
parameters for the increase of entanglement and critical
temperature. Furthermore, we analyze the properties of x-component
parameters for the control of entanglement. In the ground state,
increasing $D_{x}$ can decrease the critical value $b_{xc}$ to
increase to revival region, and increase the entanglement in the
revival region. The critical value $D_{xc}$ can also be decreased
by increasing $b_{x}$ and $J$, or decreasing $B_{x}$ and $\Delta$.
In the thermal state, for $T$ or $b_{x}$, increasing $D_{x}$ can
increase the revival region and the entanglement in the revival
region, for $B_{x}$, increasing $D_{x}$ can increase the critical
value $B_{xc}$ to largen the region of high entanglement. In
addition, decreasing $B_{x}$ can also increase the entanglement
and the revival region. Small $b_{x}$ possesses some similar
properties to $D_{x}$, and with the increase of $b_{x}$, the
entanglement also has a revival phenomenon so that the
entanglement can exist at a higher temperature for larger $b_{x}$.
Our results imply that more efficient control parameters can be
gotten by changing the direction of parameters, and more
entanglement and higher critical temperature can be obtained by
adjusting the values of these parameters. Thus, by using the state
with more entanglement as the quantum channel, the ultimate
fidelity will be increased in quantum communication. In addition,
in practice the need for low temperature enhances the complexity
and difficulty of constructing a quantum computer, however by
increasing the critical temperature the entanglement can exist at
a higher temperature, this will reduce the technical difficulties
for the realization of a quantum computer.

\begin{acknowledgments}
This work is supported by National Natural Science Foundation of
China (NSFC) under Grant Nos: 60678022 and 10704001, the
Specialized Research Fund for the Doctoral Program of Higher
Education under Grant No. 20060357008, Anhui Provincial Natural
Science Foundation under Grant No. 070412060, the Key Program of
the Education Department of Anhui Province under Grant No.
KJ2008A28ZC.
\end{acknowledgments}

\end{document}